# Time dependent Ginzburg-Landau theory for design of superconducting wire

Naoum Karchev
*Department of Physics, University of Sofia, 1164 Sofia, Bulgaria*

In the present paper, we study the superconducting wire. It is known from Maxwell equations that the current creates magnetic field that suppresses superconductivity and wire starts to conduct with resistance. We consider the time dependent Ginzburg-Landau theory to resolve the problem. The solutions of the system of equations show that applied electric field, perpendicular to the axis of the wire and to the magnetic field restores superconductivity and wire starts to conduct without resistivity. We can increase the current to a new suppression of superconductivity and to restore superconductivity increasing the applied electric field. The aforementioned results permit us to conclude that the superconducting wire can transport a very strong current if the novel design discussed in the paper is applied.

## I. INTRODUCTION

Superconductivity is the property of certain materials to conduct direct current (DC) electricity without energy loss when they are cooled below a critical temperature (referred to as Tc). The superconducting current does not transfer heat, it is not accompanied by energy dissipation and exists in thermodynamically balanced systems. Normal current is associated with the release of Joule heat.

The application of superconductivity in technology attracts a huge interest. Superconducting wires are electric wires made of superconductivity material. When cooled below transition temperature they have zero electrical resistance. Superconducting cable can transport electric power at an effective current density of over $100 A/mm^2$ which is more than 100 times that of copper cable[1]. The high power density and electrical efficiency of superconductor wire results in compact, powerful devices and systems. Usually, superconductor such as niobium titanium alloy [2, 3], $B-$based [4] and high-temperature superconductors such as YBCO [5] are used.

We consider the design of superconducting wires. Their performance is fundamentally limited by the magnetic field and their contemporary design relies on a high critical magnetic field of the superconductors, which allows for the transport of a stronger current.

To resolve the problem we propose a fundamentally different design for superconductor wire and consider the time dependent Ginzburg-Landau theory for the wire. The generalized Maxwell system of equations [6] is the main object of study. The solution shows that applied electric field increases Bose condensation of Cooper pairs and enhances superconductivity. Inspired by the above result, we think that by placing the wire in an electric field, oriented in a suitable way, the superconductivity will be restored and superconducting wire will transport a very strong current without resistivity..

We consider Ginzburg-Landau theory of superconductivity[7]. Most appropriate to study the superconducting wire is the time-dependent Ginzburg-Landau (TDGL) theory with field-theoretical action [8–10].

$$\begin{aligned} S &= \int d^4x \left[ -\frac{1}{4} \left( \partial_\lambda A_\nu - \partial_\nu A_\lambda \right) \left( \partial^\lambda A^\nu - \partial^\nu A^\lambda \right) \right. \\ &+ \frac{1}{D} \psi^* \left( i\partial_t - e^* \varphi \right) \psi \\ &- \frac{1}{2m^*} \left( \partial_k - ie^* A_k \right) \psi^* \left( \partial_k + ie^* A_k \right) \psi \\ &+ \left. \alpha \psi^* \psi - \frac{g}{2} \left( \psi^* \psi \right)^2 \right], \end{aligned} \quad (1)$$

written in terms of gauge four-vector electromagnetic potential "$A$" and complex scalar field "$\psi$"-the superconducting order parameter. The constant $D$ is the normal-state diffusion, $(e^*, m^*)$ are effective charge and mass of superconducting quasi-particles and $\varphi = vA_0$ is the electric scalar potential. We use the notations $v^{-2} = \mu\varepsilon$, where $\mu$ is the magnetic permeability and $\varepsilon$ is the electric permittivity of the superconductor. The parameter $\alpha$ is a function of the temperature $T$

$$\alpha = \alpha_0(T_c - T), \quad (2)$$

where $T_c$ is the superconducting critical temperature and $\alpha_0$ is a positive constant. In superconducting phase $\alpha$ is positive.

The action (1) is invariant under $U(1)$ gauge transformations. We represent the order parameter in the form $\psi(x) = \rho(x) \exp[ie^*\theta(x)]$, where $\rho(x) = |\psi(x)|$ is the gauge invariant local density of Cooper pairs and introduce the gauge invariant vector $Q_k = \partial_k\theta + A_k$, scalar $U = \partial_t\theta + \varphi$ and antisymmetric $F_{\lambda\nu} = \partial_\lambda A_\nu - \partial_\nu A_\lambda$ fields.

The system of equations which describes the electro-

dynamics of s-wave superconductors is [11]:

$$\vec{\nabla} \times \mathbf{B} = \mu\varepsilon \frac{\partial \mathbf{E}}{\partial t} - \frac{e^{*2}}{m^*}\rho^2 \mathbf{Q} \quad (3)$$

$$\vec{\nabla} \times \mathbf{Q} = \mathbf{B} \quad (4)$$

$$\vec{\nabla} \cdot \mathbf{E} = \frac{\mu\varepsilon e^*}{D}\rho^2 \quad (5)$$

$$\vec{\nabla} U + \frac{\partial \mathbf{Q}}{\partial t} = -\mathbf{E} \quad (6)$$

$$\frac{1}{2m^*}\Delta\rho + \alpha\rho - g\rho^3 - \frac{e^*}{D}\rho U - \frac{e^{*2}}{2m^*}\rho \mathbf{Q}^2 = 0. \quad (7)$$

where the electric $\mathbf{E}$ and magnetic $\mathbf{B}$ fields are introduced in a standard way, $(F_{01}, F_{02}, F_{03}) = \mathbf{E}/v$, $(F_{32}, F_{13}, F_{21}) = \mathbf{B}$ and $(Q^0, Q^1, Q^2, Q^3) = (U/v, \mathbf{Q})$. We are interested in the system of equations for time-independent fields An important characteristic of superconductivity is the Cooper pair density $\rho$ as a function of coordinates. A wire along the z-axis is invariant under rotations about this axis and we introduce cylindrical coordinates $(r, \phi, z)$. Then $\rho$ does not depend on $\phi$. We are considering a wire with infinite length, therefor $\rho$ does not depend on $z$ either. Without magnetic and electric fields $\mathbf{B} = \mathbf{E} = 0$, and $(\mathbf{Q} = 0, U = 0)$, we write an equation for the dimensionless function $f(r) = \frac{\rho}{\rho_0}$, where $\rho_0 = \sqrt{\frac{\alpha}{g}}$:

$$\delta \frac{1}{l}\frac{d}{dl}(l\frac{df}{dl}) + f - f^3 = 0. \quad (8)$$

We have introduced the dimensionless distance $l = \frac{r}{r_0}$ with $r_0$ the radius of the wire. The solutions of equation (8) depend on the parameter

$$\delta = \frac{2\zeta_{GL}^2}{r_0^2}, \quad (9)$$

where $\zeta_{GL}$ is the Ginzburg-Landau penetration depth. The solutions for different $\delta$ are depicted in figure (1). The curves in the figure show that the density of Cooper pairs is smaller when the radius $r_0$ of the wire is small ($\delta$ is large).

The curves in the figure are reference for the following analysis.

## II. INFLUENCE OF MAGNETIC FIELD

When a superconducting current flows, a magnetic field is created. It circulates in a closed circle around the wire. In cylindrical coordinate system only $B_\phi$ component of the magnetic field is non zero. From equation (4) follows that only z component of the vector potential $Q_z$ is nonzero. For convenience we introduce the dimensionless functions

$$B(r) = \frac{B(r)_\phi}{B_0}$$
$$Q(r) = -\frac{r_0}{\zeta^2 B_0}Q_z(r), \quad (10)$$

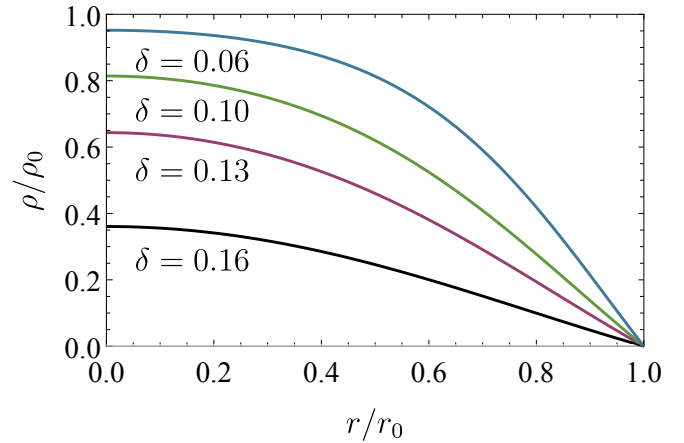

FIG. 1: The dimensionless functions $\rho/\rho_0$ of a dimensionless distance $l = r/r_0$, are plotted for different $\delta$. The maximum of the density $\rho/\rho_0$ increases when $\delta$ decreases i.e the radius of the wire increases.

where $B_0$ is proportional to maximum of the magnetic field with coefficient depending on $\delta$. The system of equation for $f(l), B(l)$ and $Q(l)$ is

$$\delta\frac{dQ}{dl} = B \quad (11)$$

$$\frac{1}{l}\frac{d}{dl}(lB) = \frac{1}{\kappa^2}f^2 Q \quad (12)$$

$$\delta\frac{1}{l}\frac{d}{dl}(l\frac{df}{dl}) + f - f^3 - \gamma\delta f Q^2 = 0. \quad (13)$$

where

$$\kappa = \sqrt{\frac{2m^* g}{e^{*2}}} \quad (14)$$

is the Ginzburg-Landau parameter. It does not depend on temperature and characterizes the superconductor. In all calculations we set $\kappa = 0.5$.

The parameter $\gamma$ is proportional to $B_0^2$

$$\gamma = \frac{e^{*2}}{4m^{*2}\alpha^2}B_0^2 \quad (15)$$

It is a measure of the magnetic field generated when a supercurrent flows.

The density of Cooper pairs for different values of $\delta$ and $\gamma$ are depicted in figure (2)

The upper values of $\gamma$ in figure are reference, equal for all $\delta$. The bottom values are the critical one, different for different $\delta$. When $\gamma$ is larger then critical one the superconductivity disappears. The most important fact is that critical value of $\gamma$ is larger when $\delta$ is larger (Fig 2a), i.e. wires with smaller radius are capable to transport stronger current. This is the reason why in the contemporary design the superconducting wires are almost always multi filament arrays in which individual filaments have a diameter $< 30\mu m$ [12].

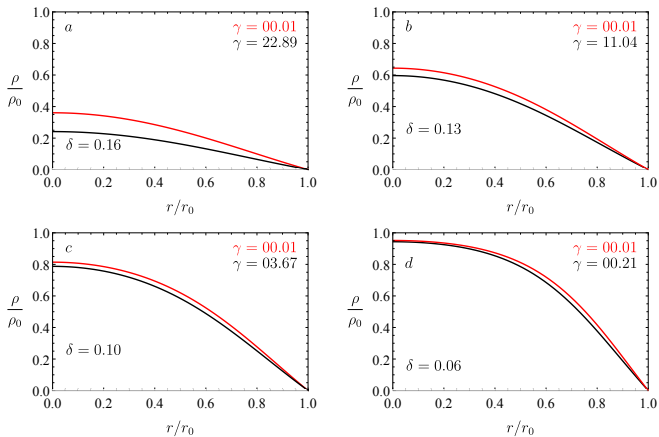

FIG. 2: The dimensionless functions $\rho/\rho_0$ of a dimensionless distance $l = r/r_0$, are plotted for different values of $\delta$ and $\gamma$. The upper values of $\gamma$ are equal for all $\delta$. The bottom values are the critical one, different for different $\delta$. When $\gamma$ is larger then critical one the superconductivity disappears. The main result is that critical value of $\gamma$ is larger when $\delta$ is larger (Fig a), i.e. wires with smaller radius are most appropriate to transport stronger current. Decrease of $\delta$ decreases the current above which the superconductivity is suppressed and the transport is with resistance.

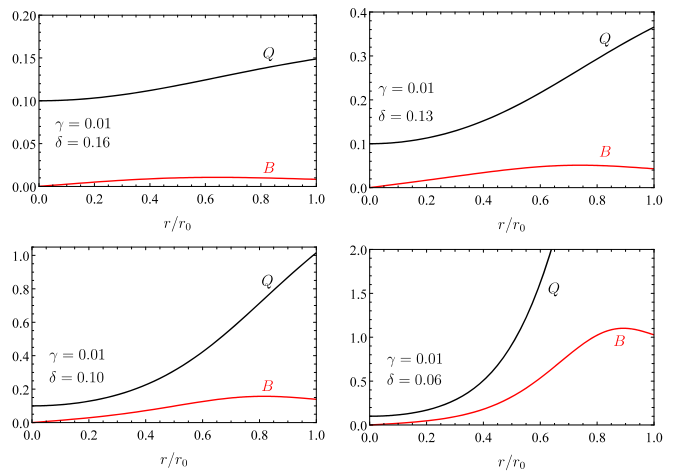

FIG. 3: The dimensionless functions $Q$ and $B$ of a dimensionless distance $l = r/r_0$, are plotted for different values of $\delta$ and $\gamma = 0.01$. For very thin conductors ($\delta = 0.16$) the London penetration depth is almost equal to $r_0$. When ($\delta = 0.06$) the magnetic field is within a thin layer of the surface. The current increase follows the increase of magnetic field and for large $r_0$ wires current flows near the surface within London penetration dept.

## III. CURRENT AND MAGNETIC FIELD IN WIRE

The current flow generates a magnetic field. In type I superconductors, it cannot exist in the interior. The wires have cylindrical geometry and only $B_\phi$ component of the magnetic field is nonzero. We calculate the dimensionless function (10). The solutions of the system of equations (11, 12, 13) demonstrate this phenomenon. Magnetic field as a function of dimensionless distance are depicted in Fig. (3).

The gauge-invariant vector field **Q** takes part in (Eq 4) as a magnetic vector potential, while in (Eq 3) is a current. This dual contribution of the vector field is the basis of electrodynamics of superconductors. In the case of superconducting wire only "z" component of the magnetic vector is nonzero. We calculate the dimensionless function $Q$ (10). The result is depicted in Fig. (3)

London penetration depth $\lambda_L$ characterizes the distance through which the weak magnetic field penetrates into the superconductors. Deep inside a superconductor, the magnetic field is zero-Meissner effect. Figure (3) shows that when the radius of conductor is small, $\delta = 0.16$ the London penetration depth is approximately equal to it. When the radius increases, the difference becomes visible and for a large radius the penetration depth is small. The magnetic field circulates in a circle around the wire is limited within London penetration depth.

The current flows along wire. Only "z" component of **Q** is nonzero. The increase of $Q$ as a function of dimensionless distance $r/r_0$ follows the increase of the magnetic field $B$. When $\delta = 0.06$ (the radius $r_0$ is large) the current flows near the surface within London penetration depth.

The current and magnetic field in the wires are discussed in [15] and many years earlier [16].

## IV. APPLIED ELECTRIC FIELD: NOVEL DESIGN OF SUPERCONDUCTING WIRE

The magnetic field is inevitable in superconducting wires. Many methods are proposed to resolve the problem and among the most popular is the aforementioned [12].

The applied electric field increases the Bose condensation of Cooper pairs and enhances superconductivity[6]. To follow this phenomenon in the case of superconducting wire we consider an applied electric field perpendicular to the axis of the wire and the magnetic field. This means that only $E_r$ component of the electric field is nonzero. For convenience we introduce the dimensionless functions

$$E(r) = \frac{E_r}{E_0} \quad (16)$$

$$q(r) = \frac{U(l)}{r_0 E_0}, \quad (17)$$

where $E_0$ is the applied electric field. The system of

equations for $f(l), B(l), E(l), Q(l)$ and $q(l)$ is

$$\delta \frac{dQ}{dl} = B \tag{18}$$

$$\frac{1}{l}\frac{d}{dl}(lB) = \frac{1}{\kappa^2}f^2 Q \tag{19}$$

$$\frac{dq}{dl} = -E \tag{20}$$

$$\frac{1}{l}\frac{d}{dl}(lE) = \frac{\lambda}{\sqrt{\delta}}f^2 \tag{21}$$

$$\delta\frac{1}{l}\frac{d}{dl}(l\frac{df}{dl}) + f - f^3 - \frac{\beta}{\sqrt{\delta}}fq - \gamma\delta f Q^2 = 0, \tag{22}$$

where $\beta$ is proportional to the applied electric field. The parameter $\lambda$ depends on the temperature and characterizes superconductors

$$\beta = \frac{\sqrt{2}\zeta_{GL}e^*}{\alpha D}E_0. \tag{23}$$

$$\lambda = \frac{\sqrt{2}\zeta_{GL}e^*\mu\epsilon}{gD}\alpha. \tag{24}$$

In all calculations we set $\lambda = 0.05$.

When $\delta = 0.16$ the critical value of current flow, above which there is not superconductivity, corresponds to $\gamma = 22.89$ (Fig 2a). We consider currents with $\gamma = 30.00$ and $\gamma = 40.00$. The values are well above the critical and the wires are conducting current with resistivity. If we apply electric field corresponding to $\beta = 02.78$ and $\beta = 04.3$ respectively the result (Fig 4a) shows that density of Cooper pairs $\rho$ is nonzero i.e. superconductivity is restored and the transport of the wires is without resistance.

For $\delta = 0.13$ the critical value is $\gamma = 11.04$ (Fig 2b). We consider wires with $\gamma = 15$ or $\gamma = 20$, values well above the critical. To restore superconductivity, respectively conductivity without resistivity, we apply an electric field with $\beta = 01.9$ or $\beta = 04.3$.

The parameter $\gamma = 03.67$ (Fig 2c) corresponds to the critical current of superconducting wire if $\delta = 0.10$. We consider wires which are not superconductors with $\gamma = 05.00$ or $\gamma = 10.00$. To restore the superconductivity we have to apply electric field with $\beta = 01.89$ or $\beta = 03.15$ respectively.

Finally, the wire with maximum radius $\delta = 0.06$ has a minimum critical current with $\gamma = 00.21$. Strong currents corresponding to $\gamma = 00.40$ or $\gamma = 01.00$ destroy superconductivity. To restore conductivity without resistivity we have to apply not so strong electric field $\beta = 00.33$ or $\beta = 01.03$

The results permit to formulate a novel design for superconducting wires. When the current is strong and the wire conducts with resistivity, one hast to apply an electric field, perpendicular to the wire's direction and the emergent magnetic field, and strong enough to restore superconductivity i.e conductivity without resistance.

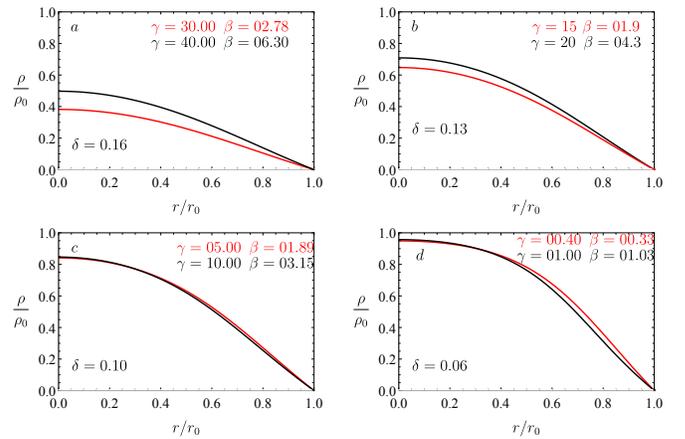

FIG. 4: The dimensionless functions $\rho/\rho_0$ of a dimensionless distance $l = r/r_0$, are plotted for different values of $\delta$, $\gamma$ and $\beta$. In all figures $\gamma$ corresponds to currents stronger then critical for relevant $\delta$, so that wire conducts with resistivity. The calculations show that one has to apply different electric fields (different $\beta$) to restore superconductivity :a) $\delta = 0.16$-$\beta = 02.78$ when $\gamma = 30.00$, $\beta = 06.30$ when $\gamma = 40.00$ b) $\delta = 0.13$-$\beta = 01.9$ when $\gamma = 15$, $\beta = 04.3$ when $\gamma = 20.00$, c) $\delta = 0.10$-$\beta = 01.89$ when $\gamma = 05.00$, $\beta = 03.15$ when $\gamma = 10.00$ and d) $\delta = 0.06$-$\beta = 00.33$ when $\gamma = 00.40$, $\beta = 01.03$ when $\gamma = 01.00$

## V. SUMMARY

In conclusion, we developed a novel design for a superconducting wire. It is based on the fact that applied electrician field increases Bose condensation of Cooper pairs and enhanced superconductivity [6]. To illustrate this, we consider time-dependent Ginzburg-Landau (TDGL) theory of superconductivity. The solutions show that superconductivity depends on the geometry (radius) of the wire. This is true when neither magnetic nor electric fields are applied. This is true when magnetic field is applied. Finally, it is true when magnetic and electric fields are applied. The phenomenon permits to make many comments concerning application of the wires with design reported in the paper.

There are principle differences between reported wires and the contemporary. What is new in the design reported in the present paper is that the conductivity is controlled by an applied electric field. For example:

-In ordinary superconducting wires, it is important that the superconductors have a strong critical magnetic field. Any superconductor can be used in the reported conductors. The weak critical magnetic field can be compensated by applying a stronger electric field.

-Wires with smaller radius are capable to transport stronger current, as proved in the section II. This fundamentally limits ordinary superconducting wires. The novel design resolve the problem. If radius of wire is small ($\delta = 0.16$) (Fig 4a) it can transport a strong current if strong electric field is applied. If the radius of the wire is large ($\delta = 0.06$) (Fig 4d) it transport relatively weak current. We can increase the current to suppression of superconductivity and to restore superconductivity increasing the applied electric field. In this way, wires with small or large radius can serve equally

All these differences demonstrate that novel superconducting wires, reported in the paper, are more effective then the old ones and therefore more appropriate for superconducting technology.We consider this very important due to the wide application: the electricity transport, transformers, generators motors and others [13].

The theory includes the parameters: $\delta$, $\gamma$, $\kappa$, $\beta$ and $\lambda$. The parameter $\delta$ is characteristic of the superconductor ($\zeta_{GL}$ Ginzburg-Landau penetration depth) and its geometry, radius. The dependence of the wire on this parameter is decisive for its design.

The current flow in superconductors coexists with a magnetic field, which reaches a maximum value at the surface of conductor. It is proportional to the current [14]. We introduce a parameter $B_0$ which is characteristic value of magnetic field proportional to maximal value with coefficient depending on $\delta$. Parameter $\gamma$ is proportional to square of $B_0^2$, therefor it is proportional to the square of current. This relationship shows the importance of the parameter for estimating the superconducting wire.

Ginzburg-Landau parameter $\kappa$ divides superconductors on type-I and type-II. We consider type-I superconductors with $\kappa = 0.5$. They are considered as not suitable because of low critical magnetic field. Above, we mentioned that these wires can transport strong current if strong electric field is applied. Our paper is focused on design of these wires.

Parameter $\beta$ is proportional to applied electric field. By means of the parameter we control the state of the wire: superconductor which transport without resistance or normal metal with resistivity.

There are neither theoretical nor experimental results that can clarify the role of the last parameter $\lambda$ in influence of electric field on superconductivity.

There is important limitation in our theory. If the material possesses metal superconductor transition, near the critical temperature the density of electrons is high, they screen the electric field, which can not influent the Cooper pairs. The theory can be applied at low temperature where density of electrons is small even zero. In this case the electric field penetrates in superconductor and influences on Cooper pairs.

The theory is also applicable if the compound in normal state is insulator.

## VI. ACKNOWLEDGMENTS

I would like to thank T. Vetsov for useful discussions.